\documentclass[a4paper,11pt]{article}
\pdfoutput=1
\usepackage{jinstpub}
\usepackage{epsfig}
\usepackage{textcomp}

\usepackage{lmodern}
\usepackage{float}
\usepackage{tikz}
\usepackage{url}
\usepackage{subcaption}
\title{Development and Characterization of 6-gap Bakelite Multi-gap Resistive Plate Chamber}

\author[a,b,c,1]{Rajesh Ganai,\note{Corresponding author.}}
\author[c]{Mitali Mondal,}
\author[a]{Shaifali Mehta,}
\author[c]{Zubayer~Ahammed,}
\author[c]{and Subhasis Chattopadhyay}

\affiliation[a]{GSI Helmholtzzentrum f$\ddot{u}$r Schwerionenforschung GmbH, Planckstrasse 1, 64291 Darmstadt, Germany.}
\affiliation[b]{Bose Institute, Centre for Astroparticle Physics and Space Sciences, P-1/12,CIT Road, Scheme VII-M Kolkata -700054, West Bengal, India.}
\affiliation[c]{Variable Energy Cyclotron Centre, 1/AF-Bidhan Nagar, Kolkata-700064, India.}
\emailAdd{rajesh.ganai.physics@gmail.com}

\abstract{The Multi-gap Resistive Plate Chamber (MRPC) is an advanced form of Resistive Plate Chamber (RPC) detector where the gas gap is divided into 
sub-gaps. MRPCs are known for their good time resolution and detection efficiency for charged particles. They have found suitable applications in 
several high energy physics experiments like ALICE in LHC, CERN, Geneva, Switzerland and STAR in RHIC, BNL, USA. As they have very good time resolution
and are of low cost, they can be a suitable replacement for very expensive scintillators used in Positron Emission Tomography Imaging. The MRPCs that
are being used nowadays are developed with glass electrodes. We have made an attempt to develop a 6-gap MRPC using bakelite electrodes. The outer 
electrodes are of dimensions 15 cm $\times$ 15 cm $\times$ 0.3 cm and the inner electrodes are of dimension 14 cm $\times$ 14 cm $\times$ 0.05 cm. 
The glossy finished electrode surfaces have not been treated with any oil like linseed, silicone for smoothness. The performance of the
detector has been studied measuring the efficiency, noise rate and time resolution with cosmic rays. This effort is towards the development of a 
prototype for Positron Emission Tomography with the Time-Of-Flight technique using MRPCs. Details of the development procedure and performance 
studies have been presented here.}
\keywords{Gaseous detectors, Resistive-plate chambers, Multi-gap Resistive plate Chamber, Bakelite and Time resolution}
\begin{document}
\maketitle
\flushbottom
%\setpagewiselinenumbers

\section{Introduction}
\begin{figure}[!htb]
\centering
  \includegraphics[height=5.0 cm, width=8.0 cm,keepaspectratio]{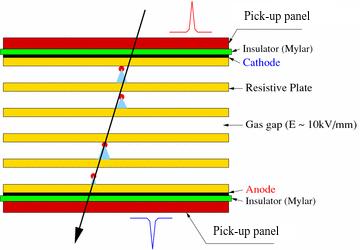}
\caption{\small \sl Schematic diagram of a MRPC\cite{mrpc_fig}.}
\label{mrpc}       % Give a unique label
\end{figure}

Multi-gap Resistive Plate Chamber (MRPC)\cite{first_mrpc} was introduced in the year 1996 with a sole purpose to obtain a much improved time 
resolution over a single gap Resistive Plate Chamber (RPC)\cite{first_RPC} without sacrificing its other good qualities like efficiency. The gas gap 
of a single gap
RPC is divided into sub-gas gaps by inserting thin (hundreds of microns thick) electrode plates inside the gas gap. The high voltages (HV)
are applied only to the external surfaces of top and bottom electrodes and the intermediate electrodes are electrically floating. Each of these 
electrodes are well separated from each other by very thin spacers. The thickness of these spacers defines each gas gap. The readout panels or 
electrodes are located outside the
stack and are well insulated from the high voltage electrodes. 
The time resolution of a single gap RPC is $\sim$1 ns\cite{rajesh_rpc} whereas the time resolution of MRPC is much better than RPC\cite{mrpc_time_reso}.
Figure ~\ref{mrpc} shows the schematic of a MRPC.

A passing charged particle may
create an avalanche in any one or more than one or all gas gaps. The avalanches in different gaps have the same time development. As the intermediate 
plates are transparent to the avalanche signals, thus induced signals on external anode and cathode are analogue sum of avalanches in all gaps. 
The time jitter in the rise time of the signal is expected to reduce due to smaller sub-gaps, resulting into less jitter in timing. This causes much better time resolution than 
single gap RPC. From the simulation studies, it has been shown that the efficiency increases and time resolution improves with the increase of 
number of gaps \cite{rpc_simulation}. If the gas gap of a single gap RPC is divided into n number of gaps to form a MRPC, then, it can been shown 
that the time resolution scales with $\sigma_t$ /$\sqrt n$ , however, the efficiency does scale with expected scaling law of 1-(1-$\epsilon$)$^{n}$, 
where $\sigma_t$  and $\epsilon$ are the
time resolution and efficiency respectively of the single gap RPC \cite{rpc_simulation}. The best time resolution obtained so far from MRPC is $\sim$15.8 ps 
\cite{mrpc_time_resolution}. Due to their excellent time resolution and detection efficiency of charged particles, several ongoing experiments 
like ALICE \cite{alice_tof_tdr}, STAR \cite{star_mrpc} use, as well as upcoming experiments like CBM \cite{cbm_tof_tdr} will use these detectors 
for timing measurements. 

The MRPCs which are being used in the experiments are made up of glass. We have given an effort to develop MRPCs with bakelite sheets. The major 
advantages of bakelite over glass are
\begin{enumerate}
\item{Bakelite RPCs can be easily operated in ``streamer mode" unlike glass RPCs reducing the number of electronics channels used and hence the overall cost in an experiment.}
\item{Unlike glass, bakelite sheets do not break easily as they have excellent mechanical strength.}
\item{Testing, handling and shifting of bakelite based modules are much more easier than the glass modules.}
\end{enumerate}
With the major disadvantages being
\begin{enumerate}
\item{The surface morphological structure of glass is smoother then bakelite.}
\item{Glass electrodes do not sag when stacked one on the other in a MRPC.}
\end{enumerate}
In the above mentioned points we have pointed out about RPCs instead of MRPCs as successful development and operation of bakelite MRPCs have not 
been reported till date. Keeping in mind the major disadvantages of bakelite over glass, we developed bakelite MRPCs as these factors can be taken 
care of. The developed MRPC is a prototype and of small dimension (15 cm $\times$ 15 cm) and placing suitable button spacers at suitable distances 
solved the sagging problem of the bakelite. After several R$\&$D's, we succeeded in getting our hands on suitable bakelite sample which did not 
require any kind of oil treatment unlike previously developed RPCs using bakelite sheets\cite{oil_rpc} while developing the detectors, hence ensuring a better surface morphological 
structure\cite{rajesh_rpc}\cite{rajesh_rpc_long_term}. The bakelite MRPC was developed keeping in mind the potential application of MRPCs in 
Medical Physics especially in Positron Emission Tomography. It is a very well known fact that MRPCs have very poor detection efficiency of 
photons at their optimized operating voltages for detecting charged particles. It has been observed that the photon detection efficiency of 
MRPCs increases as their operating voltage is increased. Increase in operating voltages leads to high gas gain mode or streamer mode of operation 
and reducing the number of electronics modules. Bakelite based RPCs are very well known to be operated in streamer mode, hence our effort for 
the development of $\bf{bakelite}$ MRPC.

\section{Development of 6-gap bakelite MRPC}
\label{devel}
\begin{figure}[htbp]
        \begin{subfigure}[b]{0.25\textwidth}
                \centering
                \includegraphics[height=3.0 cm, width=8.0 cm,keepaspectratio]{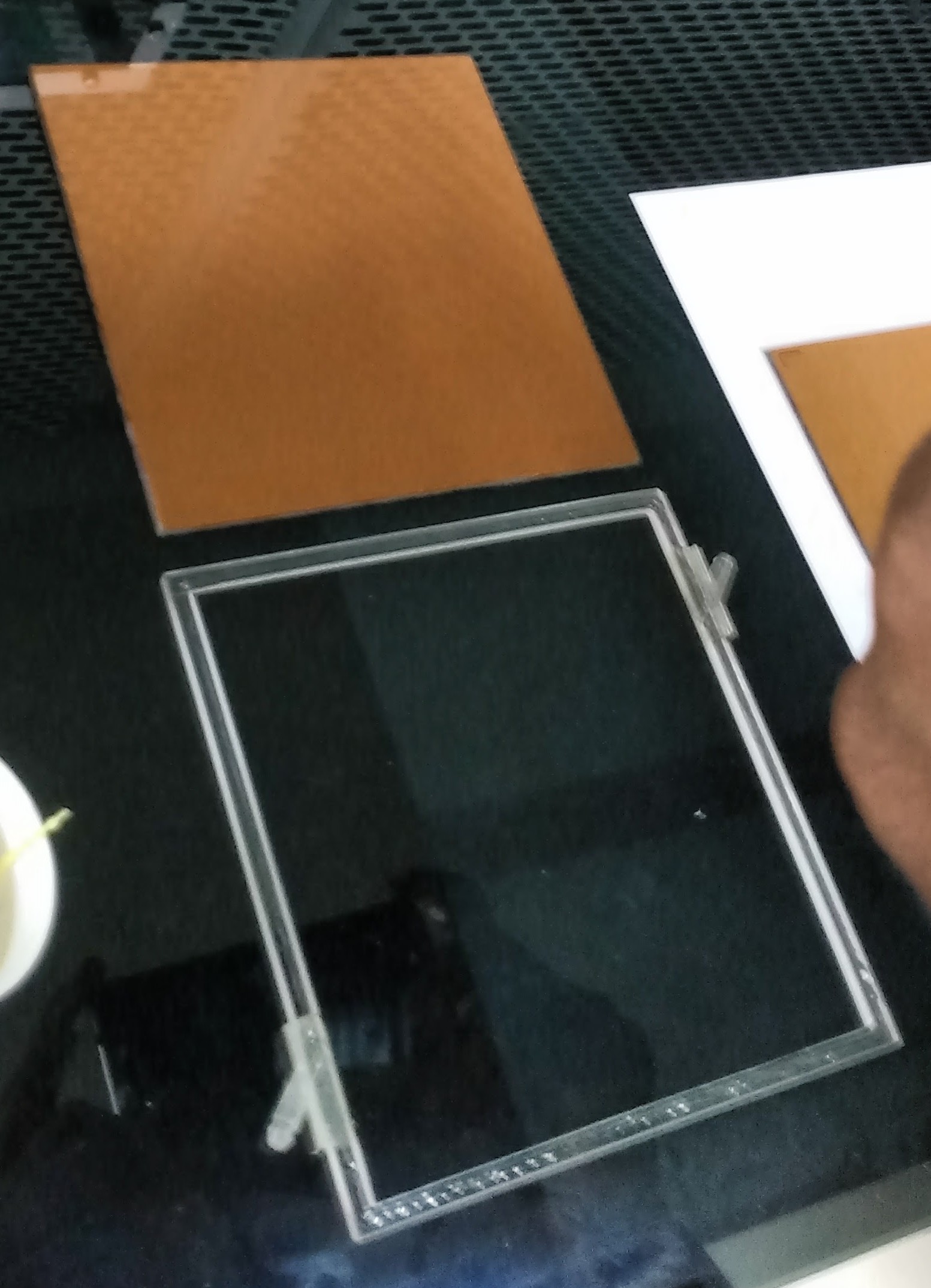}
                \caption{}
        \end{subfigure}%
        \begin{subfigure}[b]{0.25\textwidth}
                \centering
                \includegraphics[width=.85\linewidth]{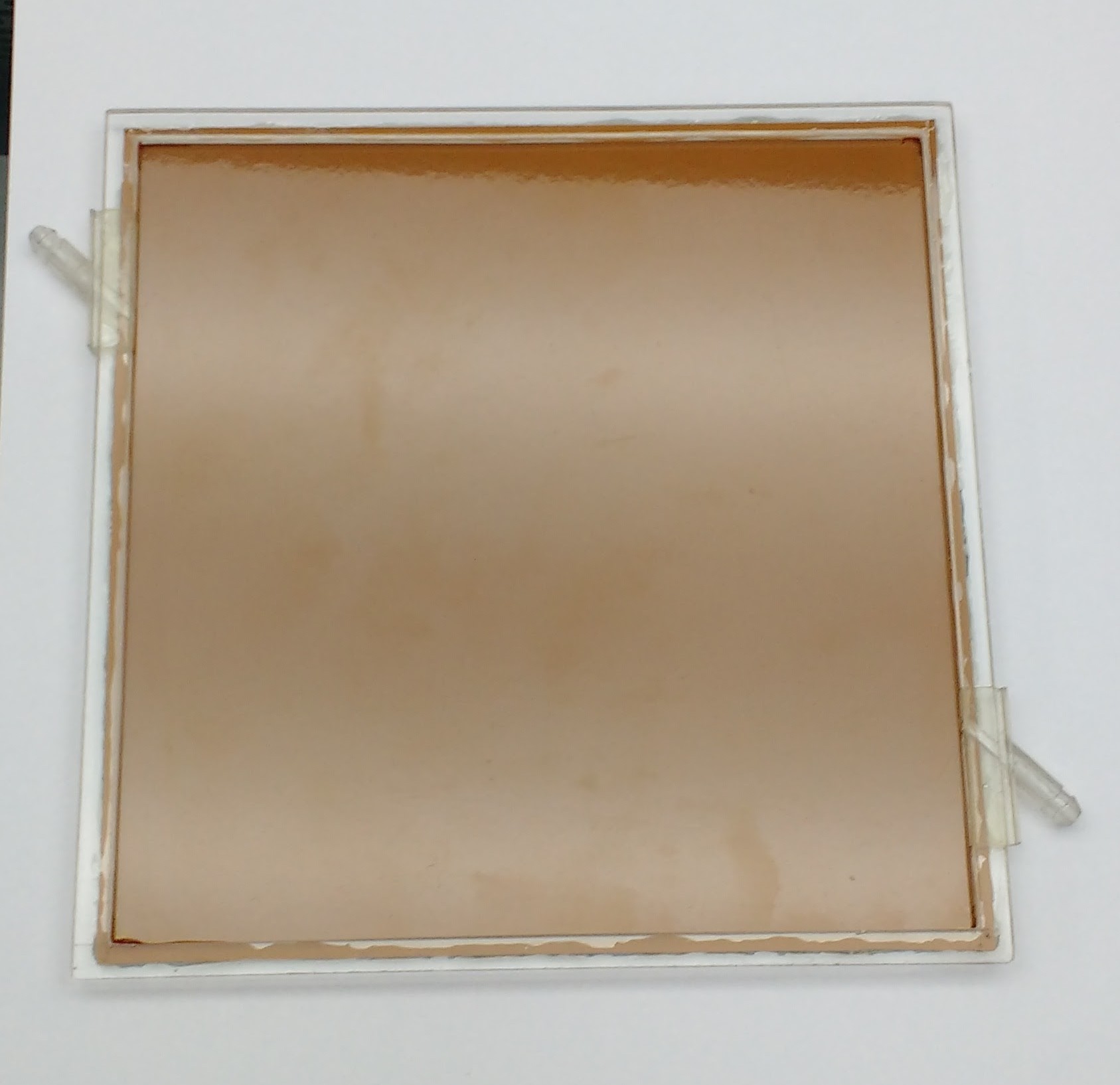}
                \caption{}
        \end{subfigure}%
        \begin{subfigure}[b]{0.25\textwidth}
                \centering
                \includegraphics[width=.90\linewidth]{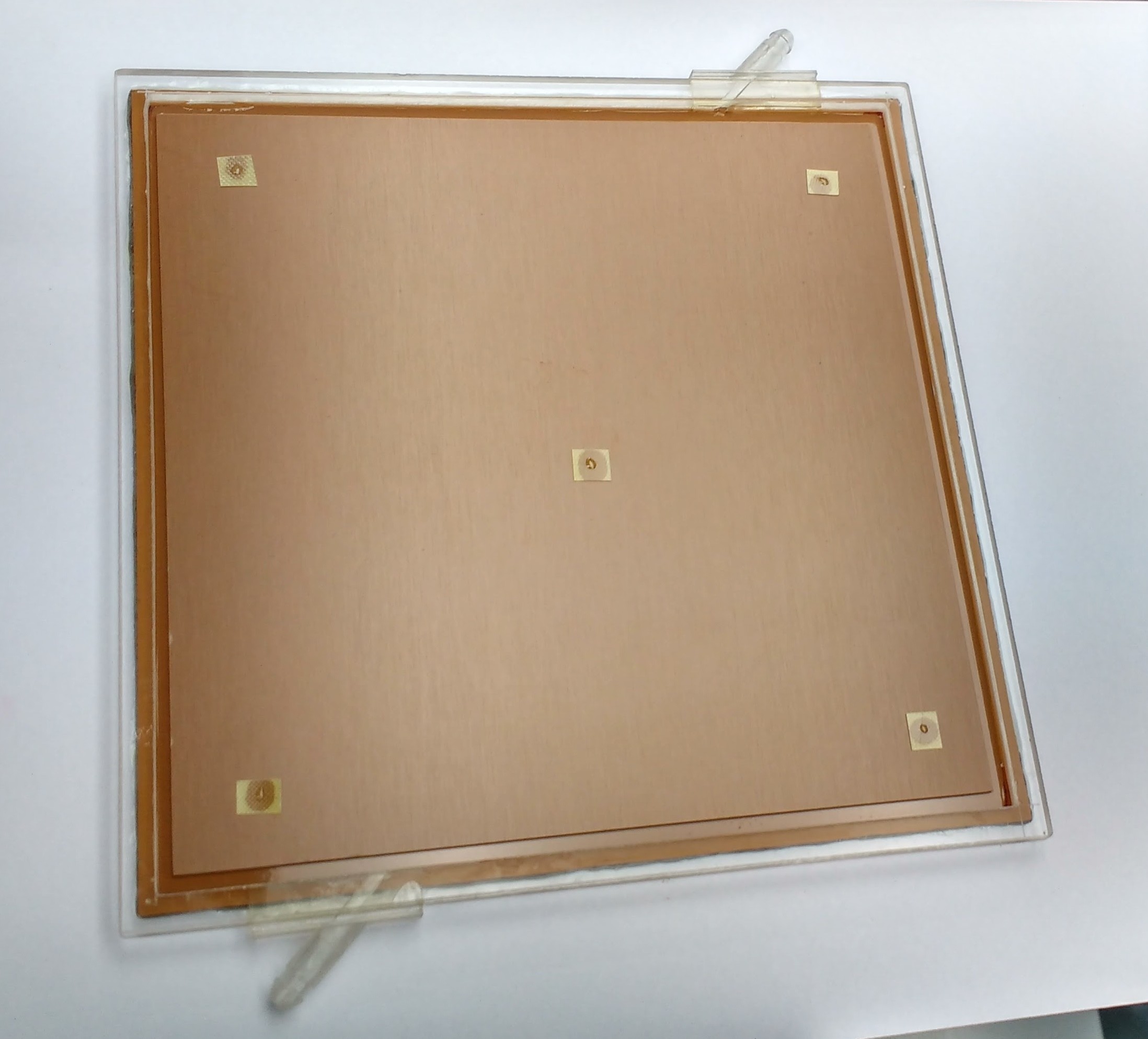}
                \caption{}
        \end{subfigure}%
        \begin{subfigure}[b]{0.25\textwidth}
                \centering
                \includegraphics[width=.95\linewidth]{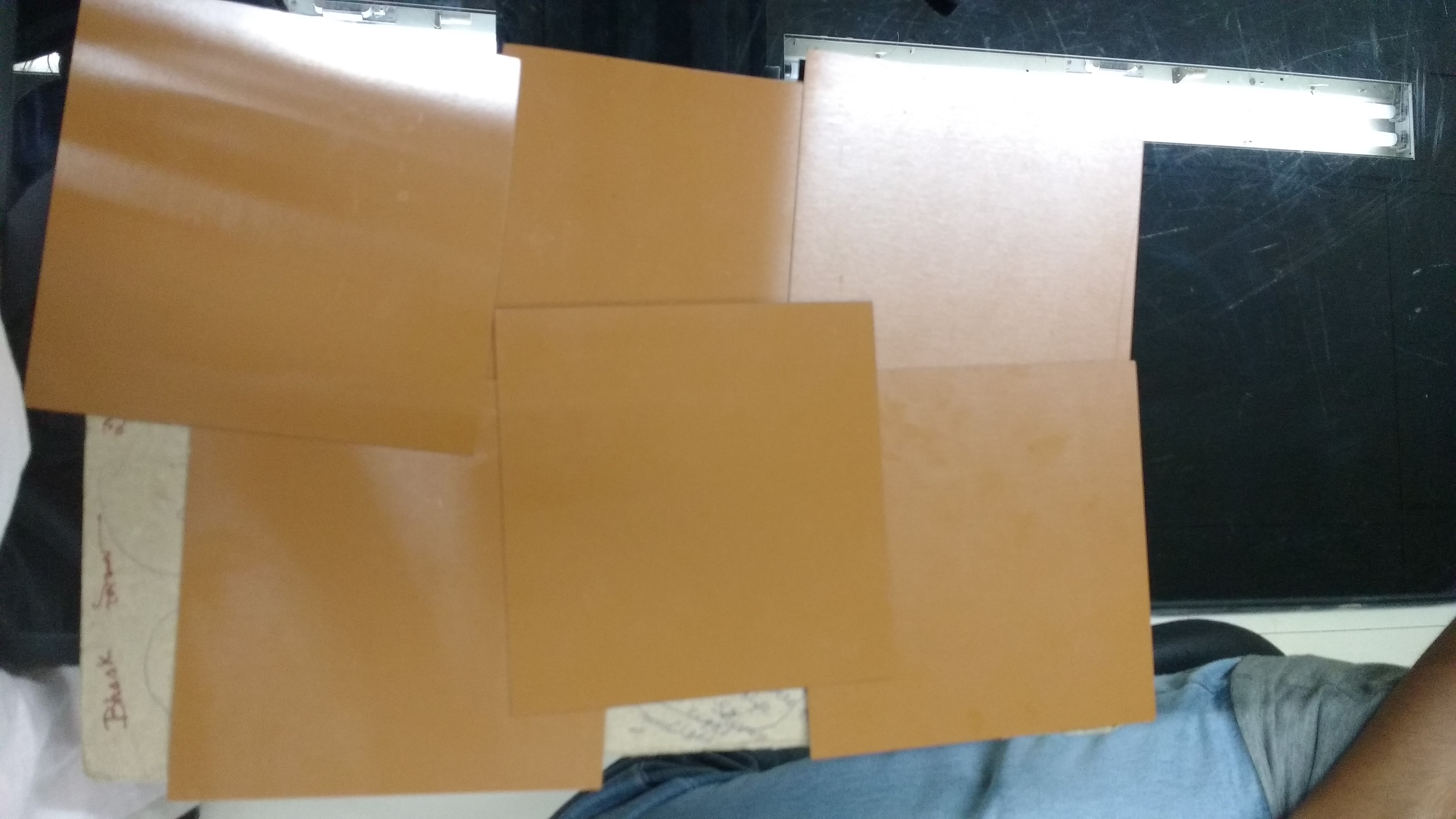}
                \caption{}
        \end{subfigure}

\par\bigskip
	\begin{subfigure}[b]{0.25\textwidth}
                \centering
                \includegraphics[width=.85\linewidth]{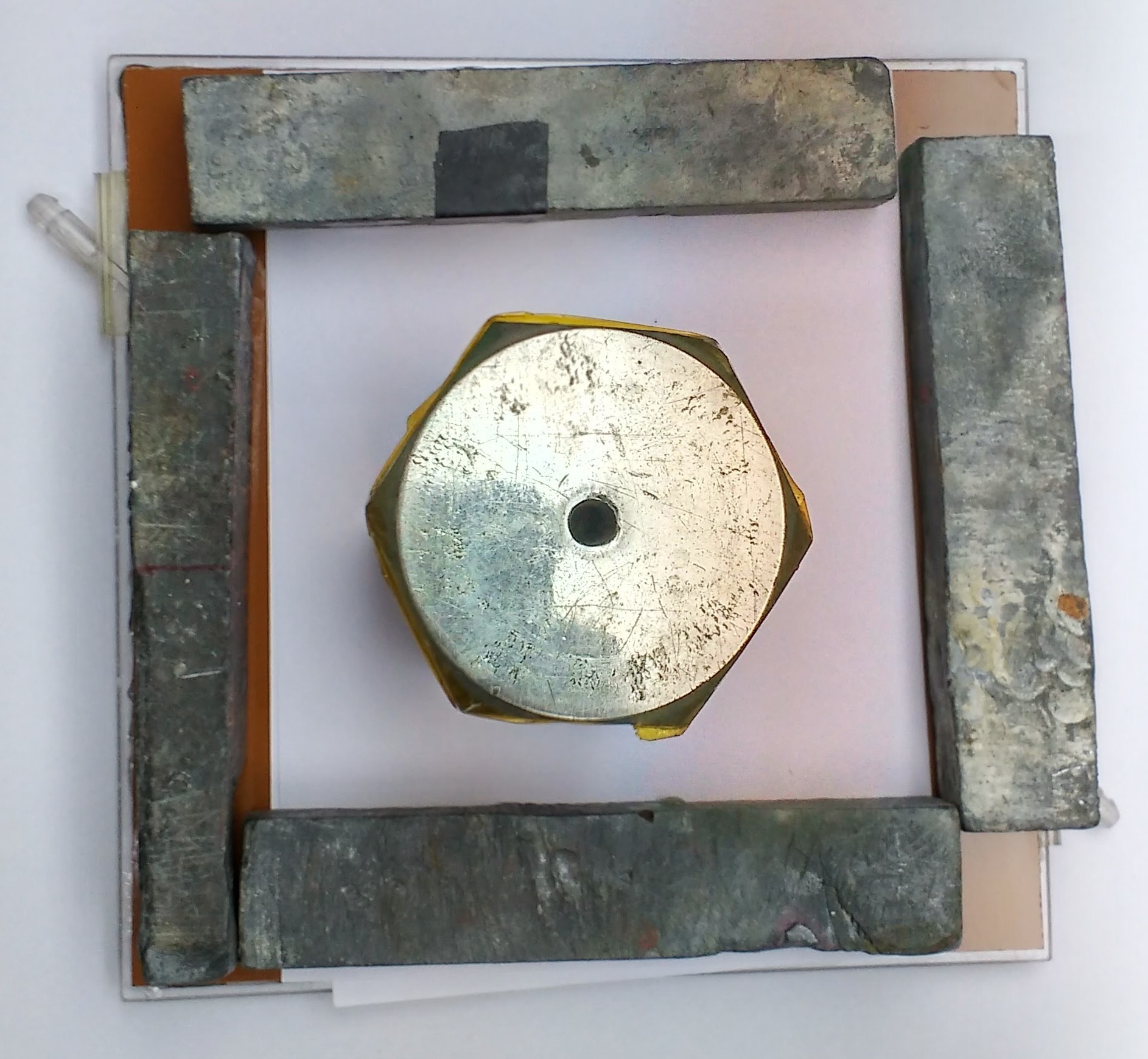}
                \caption{}
        \end{subfigure}%
        \begin{subfigure}[b]{0.25\textwidth}
                \centering
                \includegraphics[width=.65\linewidth]{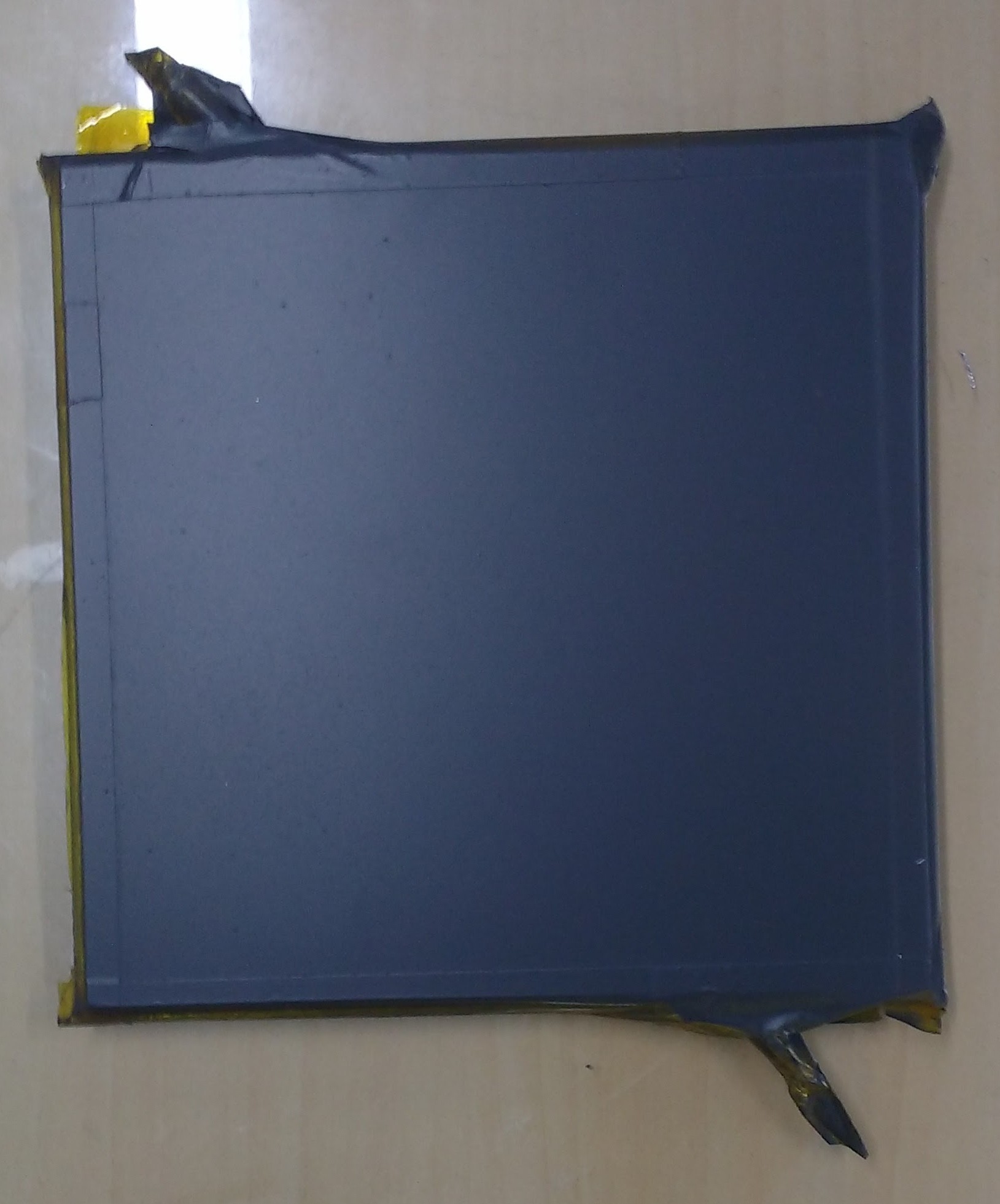}
                \caption{}
        \end{subfigure}%
        \begin{subfigure}[b]{0.25\textwidth}
                \centering
                \includegraphics[width=.90\linewidth]{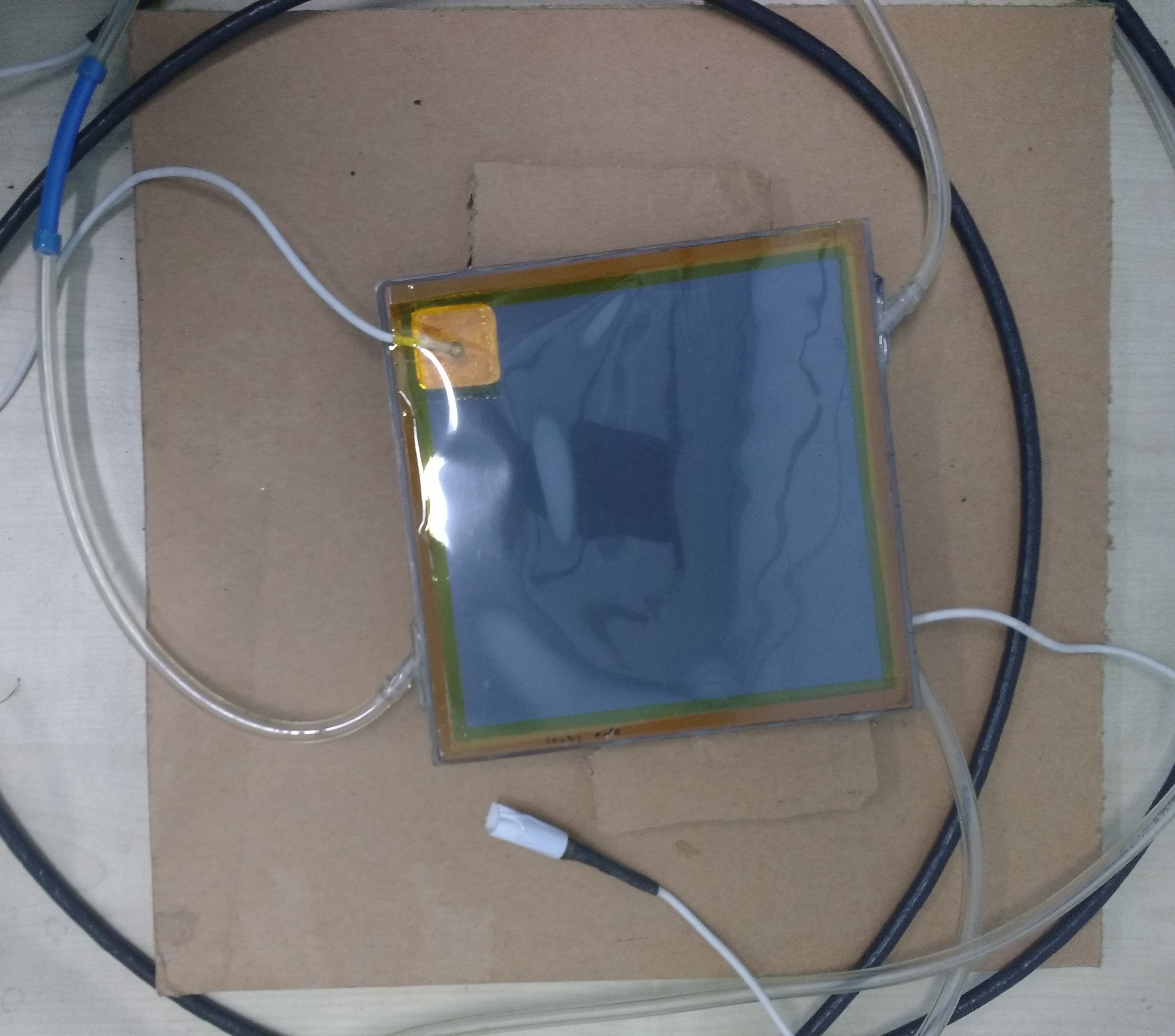}
                \caption{}
        \end{subfigure}%
        \begin{subfigure}[b]{0.25\textwidth}
                \centering
                \includegraphics[width=.95\linewidth]{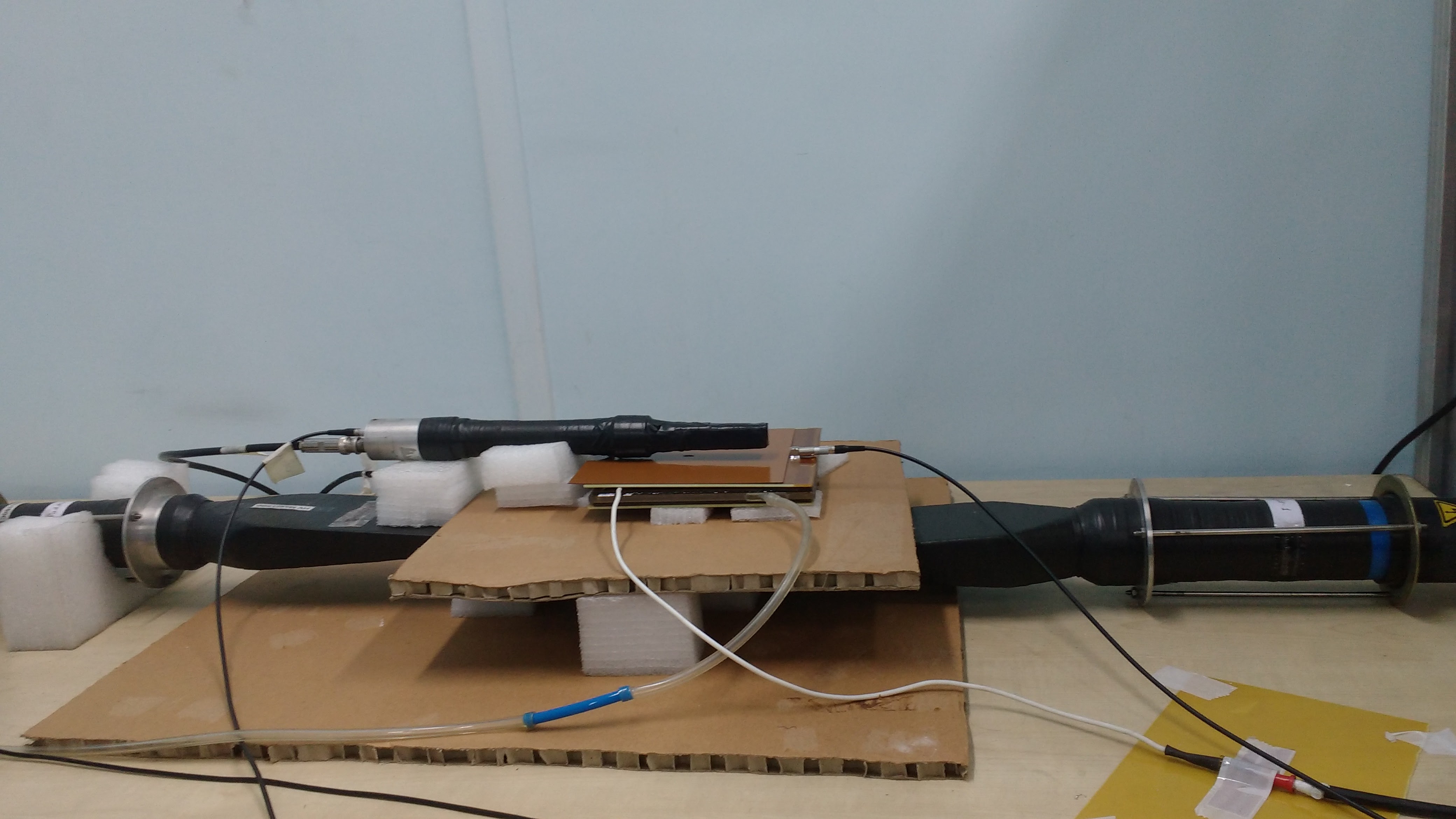}
                \caption{}
        \end{subfigure}
        \caption{\small \sl Various steps of development of bakelite MRPC (a) shows the bottom electrode plate and the side spacer with two attached gas nozzles - 
        one for gas input and another for gas output (b) shows the side spacer with the gas nozzles has been glued to the bottom electrode plate 
        (c) shows five button spacers glued on the button electrode plate (d) shows the intermediate bakelite electrodes each of $\sim$500 $\mu m$ 
        thick. These intermediate electrodes were stacked one over the other on the button electrode plate with the help of the button spacers. 
        (e) finally the whole chamber is closed by gluing the top electrode on the last intermediate electrode. Few weights have been placed on 
        the chamber to ensure that the electrode plates should cling properly with the spacers. (f) the outer surface of the top and bottom 
        electrodes was pained with semi-conducting paint. (g) the electrical and the gas connections have been done. The painted surface of the 
        electrodes were also properly insulated and isolated from the outer environment with the help of mylar sheets. (h) MRPC under test with 
        cosmic rays with one finger and two paddle scintillators.}
\label{mrpc_steps}
\end{figure}

In this section we have discussed about the development of the MRPC.
The MRPC has been developed with bakelite available in local Indian market. The details of the bakelite has been well mentioned in reference 
\cite{rajesh_rpc}. The bakelite plates (inner and outer) have not been treated with any kind of oil. The thickness of the outer and inner 
electrodes were 0.30 cm and 0.05 cm respectively. Table~\ref{mrpc_details} summarizes the details about the developed MRPC. Figure~\ref{mrpc_steps} 
shows various pictures of the development of the MRPC.
\begin{table}[htbp]
\begin{center}	
\caption{\small \sl Details of the bakelite MRPC.}
\label{tab:1}       % Give a unique label
\begin{tabular}{||l|l||}
\hline\noalign{\smallskip}
Total area of the MRPC & $\sim$ 15 cm $\times$ 15 cm   \\
\noalign{\smallskip}\hline\noalign{\smallskip}
Active area of the MRPC &  $\sim$ 14 cm $\times$ 14 cm  \\
\noalign{\smallskip}\hline\noalign{\smallskip}
Number of outer electrodes &  2  \\
\noalign{\smallskip}\hline\noalign{\smallskip}
Number of inner electrodes &  5  \\
\noalign{\smallskip}\hline\noalign{\smallskip}
Dimensions of the outer electrodes &  $\sim$ 15 cm $\times$ 15 cm $\times$ 0.30 cm  \\
\noalign{\smallskip}\hline\noalign{\smallskip}
Dimensions of the inner electrodes & $\sim$ 14 cm $\times$ 14 cm $\times$ 0.050 cm  \\
\noalign{\smallskip}\hline\noalign{\smallskip}
Thickness of each button spacer & $\sim$ 0.024 cm  \\
\noalign{\smallskip}\hline\noalign{\smallskip}
Thickness of the side spacer frame & 0.40 cm  \\
\noalign{\smallskip}\hline\noalign{\smallskip}
Total number of gas nozzles & 2  \\
\noalign{\smallskip}\hline\noalign{\smallskip}
Total number of gas gaps & 6  \\
\noalign{\smallskip}\hline\noalign{\smallskip}
Thickness of each gas gap & $\sim$ 0.025 cm  \\
\noalign{\smallskip}\hline\noalign{\smallskip}
\end{tabular}
\label{mrpc_details}
\end{center}
\end{table}
A specially made frame, as shown in Figure~\ref{mrpc_steps} (a), out of 0.4 cm thick polycarbonate sheet and two gas nozzles properly pasted on the 
frame served the purpose of the side spacer. The inner and outer length (breadth) of the frame is 14.5 cm and 15.5 cm respectively. The glue used in 
the MRPC is the same as that used in reference \cite{rajesh_rpc}. The outer surfaces of the top and bottom electrodes were spray-painted with a 
black semi-conducting paint mixed with a special dry thinner in the ratio 1:1 by volume, both manufactured by Kansai Nerolac, India.

% Define block styles

\section{Cosmic ray test results}
\label{results}
In this section we have discussed the test results (efficiency, noise rate and time resolution measurements) of the MRPC with cosmic rays at 
threshold values of 10 mV and 20 mV to the MRPC signals.

The MRPC has been tested with cosmic rays with the help of three plastic scintillators - two
paddle scintillators (20 cm $\times$ 8.5 cm) and one finger scintillator (7 cm $\times$ 1.5 cm). 
The overlap area between the scintillators has been used to obtain 
the cosmic ray efficiency for a particular set-up. The master trigger was generated from the coincidence of the three scintillator logic signals. 
The average master trigger rate was $\sim$ 0.008 Hz/$cm^2$.
A CANBERRA QUAD CFD 454 constant fraction
discriminator (CFD) has been used to digitize the signals from the scintillators and the MRPC. For timing measurements, a 16 channel PHILIPS 
SCIENTIFIC 7186 TDC module was used.
A CAMAC based data acquisition system has been used in our setup. The high voltage module used was CAEN N471A, 2 Channel 8 kV Power Supply. 
Hence, we had a limitation that we could not go beyond an applied voltage of 16 kV ($\pm$8 kV).

All the tests have been done with a gas composition of R134a:Iso-butane::85:15 by volume. A typical gas flow rate of $\sim$0.21 liters/hour 
has been maintained over the entire test period.
\subsection{Efficiency and noise rate measurements}
\begin{figure}[htbp]
\centering
  \includegraphics[height=5.0 cm, width=8.0 cm,keepaspectratio]{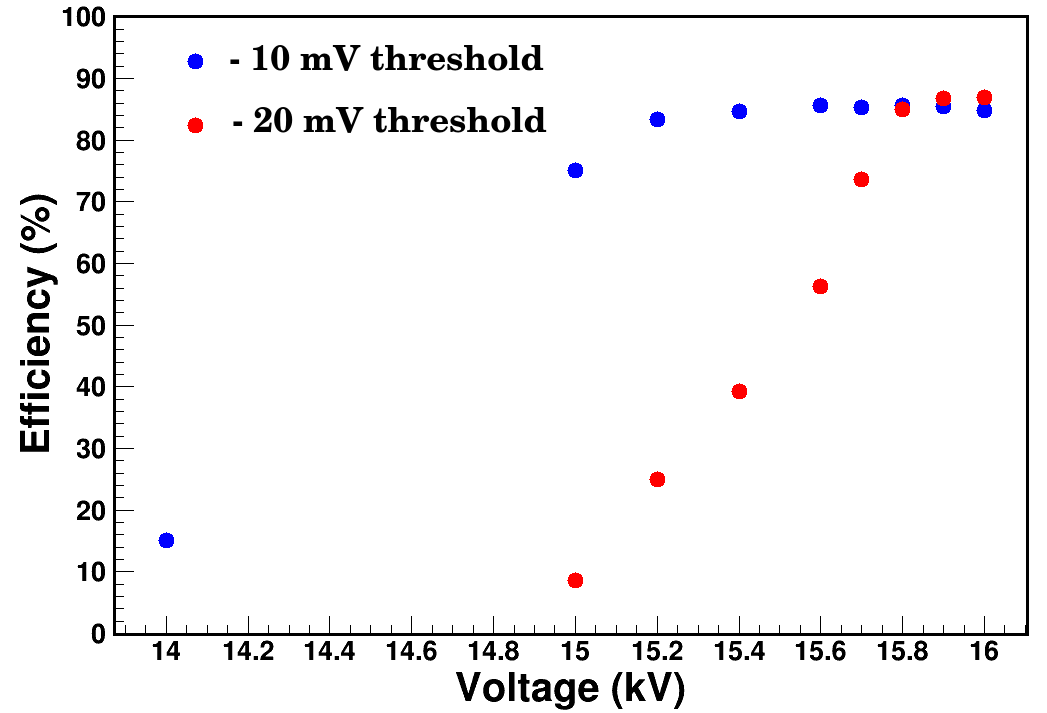}
\caption{\small \sl Variation of the efficiency of the MRPC as a function of the applied high voltage at 10 mV and 20 mV threshold. The error bars are within 
the marker size.}
\label{eff}       % Give a unique label
\end{figure}

\begin{figure}[htbp]
\centering
  \includegraphics[height=5.0 cm, width=8.0 cm,keepaspectratio]{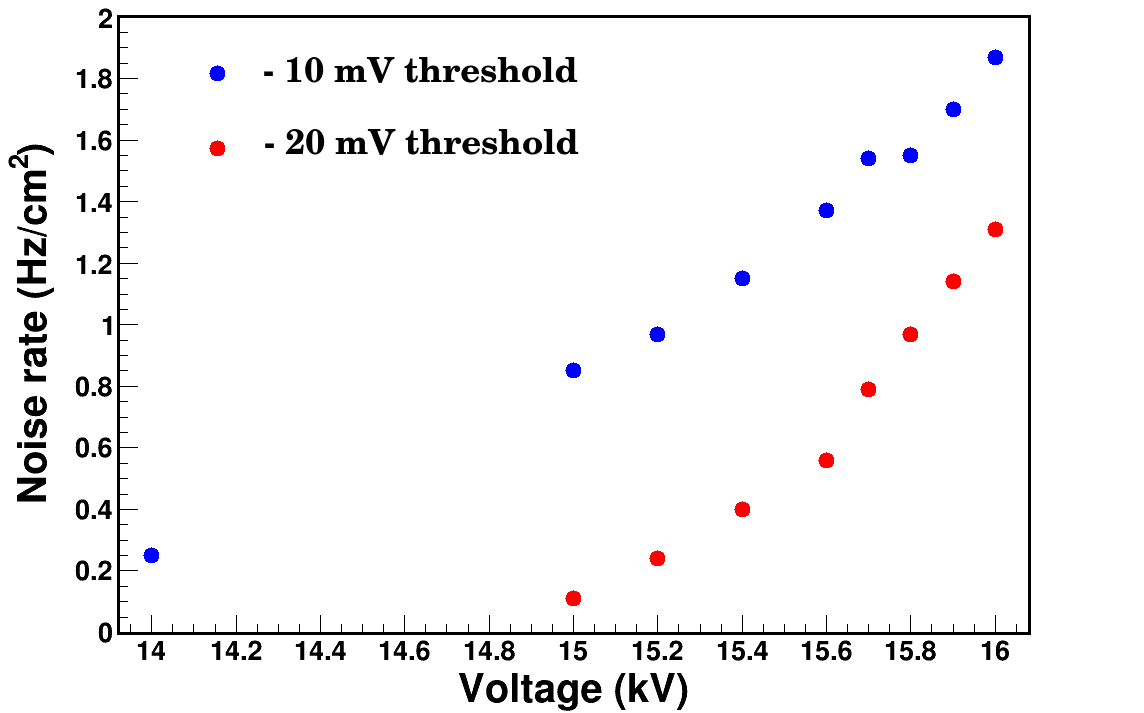}
\caption{\small \sl Variation of the noise rate of the MRPC as a function of the applied high voltage at 10 mV and 20 mV threshold. The error bars are 
within the marker size.}
\label{noise_rate}       % Give a unique label
\end{figure}

We have measured the efficiency of the MRPC with cosmic rays at two different negative thresholds to the analog signals of the detector. 
Figure ~\ref{eff} shows the variation of the efficiency as a function of the applied high voltage at 10 mV and 20 mV threshold.
A plateau of $>$85$\%$ efficiency was obtained for both the threshold values. For 10 mV threshold value, the plateau started from $\sim$15.2 kV 
whereas for 20 mV value, the same started from $\sim$15.8 kV. This behavior is expected with an increase in the threshold value. We have also 
measured the noise rate of the detector as a function of the high voltage which is shown in Figure~\ref{noise_rate}.
The noise rate values for both the threshold values followed nearly a linear behavior. It is seen from the figure that the noise rate at 10 mV and 
20 mV thresholds are $\sim$1.9 Hz/cm$^{2}$ and $\sim$1.3 Hz/cm$^{2}$ respectively at 16 kV applied voltage. As expected, the noise rate  of the 
MRPC decreased with an increase in the threshold at any particular applied voltage.

\subsection{Time resolution measurements}
\begin{figure}[htbp]
\centering
  \includegraphics[height=5.0 cm, width=8.0 cm,keepaspectratio]{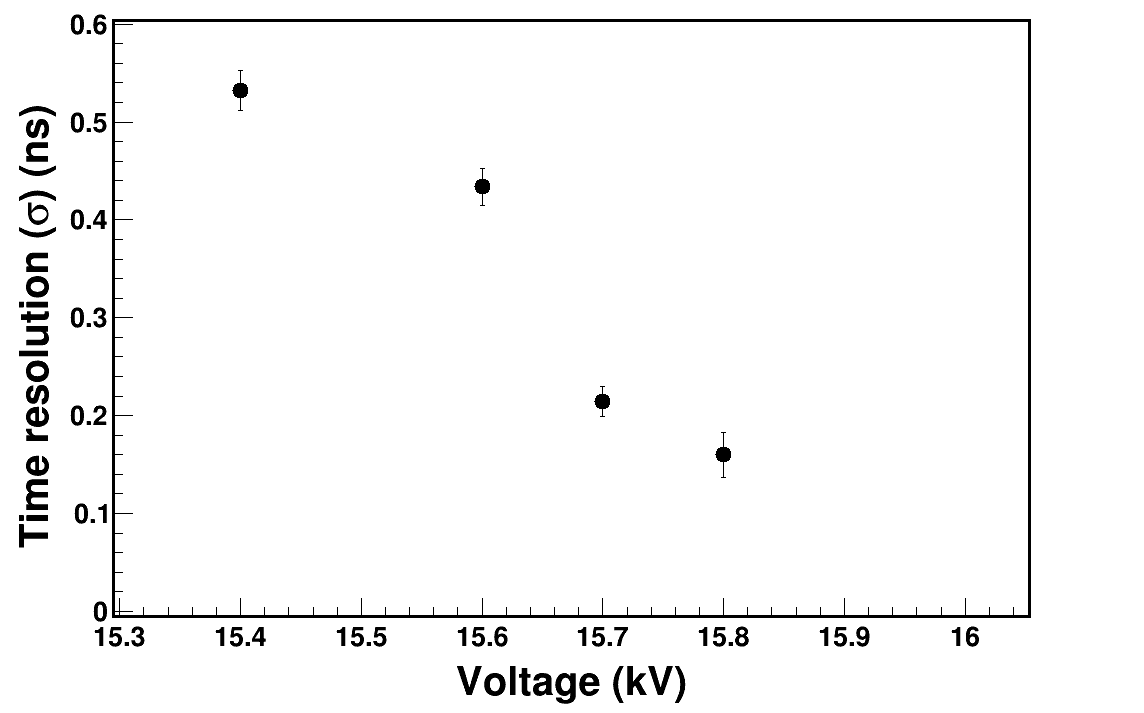}
\caption{\small \sl The scintillator corrected time resolution (ns) of the MRPC as a function of the applied high voltage at a discriminating threshold of 
10 mV.}
\label{tdc_voltage}       % Give a unique label
\end{figure}

The time resolution of the MRPC at different thresholds as well as at different applied voltages have been measured. Initially, the time spectra were recorded 
for a reasonable duration of time and later fitted with Gaussian function. The final RPC time resolution ($\sigma^{corrected}$) has been extracted 
from the Gaussian fit after removing the contribution of the three scintillators\cite{fonte_p}.
The time spectra of the MRPC have been recorded from 15.4 kV to 15.8 kV of applied high voltages at 4 different values. Figure ~\ref{tdc_voltage} 
shows the variation of the time resolution of the MRPC as a function of the applied voltage at a signal threshold of 10 mV. As seen from the figure, 
the time resolution gets better as the applied high voltage was increased and the best value of the time resolution obtained was 
$\sim$ 0.160 ns. 

\section{Summary and outlook}
\label{sum}
We have successfully developed a 6-gap bakelite MRPC. The detector has been tested for its detection efficiency and noise rate with cosmic rays at 
two different thresholds (10 mV and 20 mV) with a gas composition of R134a (85$\%$) and iso-butane (15$\%$) and typical gas flow rate of $\sim$0.21 
liters/hour. The best efficiency achieved was $>$85$\%$ at both the thresholds. The efficiency plateau started form $\sim$15.2 kV and $\sim$15.8 kV at 
10 mV and 20 mV thresholds respectively.The maximum noise rate measured were 1.9 Hz/cm$^{2}$ at 10 mV threshold and 1.3 Hz/cm$^{2}$ at 20 mv threshold. 
The best scintillator corrected time resolution measured were $\sim$160 ps at 15.8 kV at 10 mV threshold.

As future work, the chamber has to be tested with Argon based gas mixtures as well as increasing the R134a content of the gas mixture enabling us to 
reduce the optimized operating voltage. The detector has to be characterized for its photon detection efficiency. R$\&$D's have to be done to further 
improve the time resolution as well as the efficiency plateau of the MRPC.

\section{Acknowledgments}
The work is partially supported by the DAE-SRC award project fund of S. Chattopadhyay. We acknowledge the service rendered by Ganesh Das 
for fabricating the detector in several try outs.

\end{document}